\documentclass{ceurart}

\usepackage[binary-units=true]{siunitx}
\usepackage{numprint}
\usepackage{csvsimple}
\usepackage{ifthen}
\usepackage{datatool}
\usepackage{pgfplots}
\usepackage{pgfplotstable}
\usepackage{csquotes}
\usepgfplotslibrary{units}
\usepgfplotslibrary{statistics}
\usepgfplotslibrary{groupplots}
\usetikzlibrary{positioning}
\pgfplotsset{compat=1.17}
\usetikzlibrary{external}
\def\stripzero#1{\expandafter\stripzerohelp#1}
\def\stripzerohelp#1{\ifx 0#1\expandafter\stripzerohelp\else#1\fi}
\newcommand{\cor}[3]{\ifdim#3 pt<0.001 pt {$r(\pgfmathparse{int(#1-2)}\pgfmathresult)=\npdecimalsign{.}\npnoaddmissingzero\nprounddigits{2}\numprint{\stripzero{#2}}$, $p<.001$}\else {$r(\pgfmathparse{int(#1-2)}\pgfmathresult)=\npdecimalsign{.}\npnoaddmissingzero\nprounddigits{2}\numprint{\stripzero{#2}}$,  $p=\npdecimalsign{.}\npnoaddmissingzero\nprounddigits{3}\numprint{\stripzero{#3}}$}\fi{}}

\hypersetup{breaklinks=true}
\def\numCpus{54}
\def\numRuns{669}
\def\numConf{594}
\def\cpus{11th Gen Intel(R) Core(TM) i5-1135G7 @ 2.40GHz,11th Gen Intel(R) Core(TM) i7-1165G7 @ 2.80GHz,11th Gen Intel(R) Core(TM) i7-11800H @ 2.30GHz,11th Gen Intel(R) Core(TM) i7-1185G7 @ 3.00GHz,12th Gen Intel(R) Core(TM) i7-12700K,12th Gen Intel(R) Core(TM) i7-12700K w/o HT,2x AMD EPYC 7742 64-Core Processor,2x AMD EPYC 7742 64-Core Processor w/ NUMA,AMD EPYC 7601 32-Core Processor,AMD EPYC 7702 64-Core Processor,AMD EPYC 7702P 64-Core Processor (VM slice),AMD Ryzen 5 2600 Six-Core Processor,AMD Ryzen 5 3600 6-Core Processor,AMD Ryzen 5 4500U with Radeon Graphics,AMD Ryzen 5 5600X 6-Core Processor,AMD Ryzen 5 5600X 6-Core Processor (Silent Mode),AMD Ryzen 7 4800H with Radeon Graphics,AMD Ryzen 7 PRO 5850U with Radeon Graphics (Screen 1080p),AMD Ryzen 7 PRO 5850U with Radeon Graphics (Screen 2160p),AMD Ryzen 9 3900X 12-Core Processor,AMD Ryzen 9 5900X 12-Core Processor,AMD Ryzen 9 5950X 16-Core Processor (OC),AMD Ryzen Threadripper 2950X 16-Core Processor,AMD Ryzen Threadripper 2990WX 32-Core Processor,Apple M1 (using Rosetta Emulation),Apple M1 in docker (aarch64),Apple M1 Pro,Apple M1 Pro (native arm64),Fujitsu A64FX,Intel(R) Core(TM) i5-5250U CPU @ 1.60GHz,Intel(R) Core(TM) i5-5300U CPU @ 2.30GHz,Intel(R) Core(TM) i5-6200U CPU @ 2.30GHz,Intel(R) Core(TM) i5-6600K CPU @ 3.50GHz,Intel(R) Core(TM) i5-6600K CPU @ 4.50GHz,Intel(R) Core(TM) i5-7600K CPU @ 3.80GHz,Intel(R) Core(TM) i5-8259U CPU @ 2.30GHz,Intel(R) Core(TM) i7-10875H CPU @ 2.30GHz,Intel(R) Core(TM) i7-10875H CPU @ 2.30GHz w/o HT,Intel(R) Core(TM) i7-3960X CPU @ 3.30GHz,Intel(R) Core(TM) i7-4770 CPU @ 3.40GHz,Intel(R) Core(TM) i7-4790 CPU @ 3.60GHz,Intel(R) Core(TM) i7-6700 CPU @ 3.40GHz,Intel(R) Core(TM) i7-7700HQ CPU @ 2.80GHz,Intel(R) Core(TM) i7-7820HQ CPU @ 2.90GHz,Intel(R) Core(TM) i7-7920HQ CPU @ 3.10GHz,Intel(R) Core(TM) i7-8550U CPU @ 1.80GHz,Intel(R) Core(TM) i7-8559U CPU @ 2.70GHz,Intel(R) Core(TM) i7-8565U CPU @ 1.80GHz,Intel(R) Core(TM) i7-8750H CPU @ 2.20GHz,Intel(R) Core(TM) i7-9700K CPU @ 3.60GHz,Intel(R) Core(TM) i7-9700K CPU @ 3.60GHz (turbo 4.90 GHz),Intel(R) Core(TM) i7-9750H CPU @ 2.60GHz,Intel(R) Core(TM) i9-10850K CPU @ 3.60GHz,Intel(R) Core(TM) i9-10850K CPU @ 3.60GHz w/o HT,Intel(R) Core(TM) i9-10900 CPU @ 2.80GHz,Intel(R) Core(TM) i9-10900 CPU @ 2.80GHz w/o HT,Intel(R) Core(TM) i9-10910 CPU @ 3.60GHz,Intel(R) Core(TM) i9-8950HK CPU @ 2.90GHz,Intel(R) Core(TM) i9-9880H CPU @ 2.30GHz,Intel(R) Core(TM) i9-9900 CPU @ 3.10GHz,Intel(R) Core(TM) i9-9900K CPU @ 3.60GHz,Intel(R) Core(TM) i9-9980HK CPU @ 2.40GHz,Intel(R) Xeon(R) CPU E5-2699 v4 @ 2.20GHz,Intel(R) Xeon(R) Gold 6144 CPU @ 3.50GHz,Intel(R) Xeon(R) Gold 6240R CPU @ 2.40GHz,Intel(R) Xeon(R) Platinum 8360H CPU @ 3.00GHz,Intel(R) Xeon(R) Platinum 8360Y CPU @ 2.40GHz,Intel(R) Xeon(R) Platinum 8380 CPU @ 2.30GHz,Intel(R) Xeon(R) Silver 4116 CPU @ 2.10GHz,Intel(R) Xeon(R) W-2140B CPU @ 3.20GHz}
\def\bestTime{107}
\def\worstTimeH{11.2080556}
\def\bestServerTime{184}
\def\heapOutlier{0.657892}
\def\boostFreqOneTCor{\num[round-mode=places,round-precision=2]{0.5534098}}
\def\baseFreqMulTCor{\num[round-mode=places,round-precision=2]{0.3724594}}
\def\passmarkOneTRSq{0.8247232}
\def\passmarkMulTRSq{0.6784529}

\def\xdmarkOneTRSq{0.7275528}
\def\xdmarkMulTRSq{0.5845242}

\def\geekbenchOneTRSq{0.8116216}
\def\geekbenchMulTRSq{0.6969933}

\def\cinebenchOneTRSq{0.6272175}
\def\cinebenchMulTRSq{0.590943}

\def\himenoOneTRSq{0.6210778}
\def\himenoMulTRSq{0.4563649}

\def\namdMulTRSq{0.7307076}
\def\dolfynOneTRSq{0.7864744}
\def\dolfynMulTRSq{0.5242167}
\def\cpuSpecsMulTCor{0.5534098}
\def\xdmarkEightTRSq{0.7683492}
\def\xdmarkSixteenTRSq{0.6113401}
\def\modelAX{0.062931}
\def\modelB{46.7628609}
\def\modelRSq{0.8678399}
\def\modelFinalRSq{0.8442465}
\def\modelFinalTimeMae{46.5588704}
\def\effLin{0.0653066}
\def\effEnd{0.0024071}

\def\modelPrivateFinalTimeMae{37.1461619}

\begin{document}

%
% pgf preamble
\pgfplotsset{
    discard if not/.style 2 args={
        x filter/.code={
            \edef\tempa{\thisrow{#1}}
            \edef\tempb{#2}
            \ifx\tempa\tempb
            \else
                \def\pgfmathresult{inf}
            \fi
        }
    }
}
\makeatletter
\pgfplotstableset{
    discard if not/.style 2 args={
        row predicate/.code={
            \def\pgfplotstable@loc@TMPd{\pgfplotstablegetelem{##1}{#1}\of}
            \expandafter\pgfplotstable@loc@TMPd\pgfplotstablename
            \edef\tempa{\pgfplotsretval}
            \edef\tempb{#2}
            \ifx\tempa\tempb
            \else
                \pgfplotstableuserowfalse
            \fi
        }
    }
}
\makeatother

\copyrightyear{2022}
\copyrightclause{Copyright for this paper by its authors.
  Use permitted under Creative Commons License Attribution 4.0
  International (CC BY 4.0).}

\conference{PAAR'22: 8th Workshop on Practical Aspects of Automated Reasoning, August 11--12, 2022, Haifa, Israel}

\title{The Isabelle Community Benchmark}

\author{Fabian Huch}[%
orcid=0000-0002-9418-1580,
email=huch@in.tum.de,
url=https://home.in.tum.de/~huch,
]

\author{Vincent Bode}[
email=vincent.bode@tum.de,
orcid=0000-0003-2353-389X, 
url=https://home.in.tum.de/~bodev
]

\address{Technische Universität München, Boltzmannstraße 3, 85748 Garching, Germany}

% !TeX root = ../main.tex

\begin{abstract}
Choosing hardware for theorem proving is no simple task:
automated provers are highly complex and optimized programs,
often utilizing a parallel computation model,
and there is little prior research on the hardware impact on prover performance.
To alleviate the problem for Isabelle,
we initiated a community benchmark where the build time of HOL-Analysis is measured.
On \num{\numCpus} distinct CPUs,
a total of \num{\numRuns} runs with different Isabelle configurations were reported by Isabelle users.
Results range from \SI{\bestTime}{\second} to over \SI[round-mode=places,round-precision=0]{\worstTimeH}{\hour}.
We found that current consumer CPUs performed best,
with an optimal number of \numrange{8}{16} threads,
largely independent of heap memory.
As for hardware parameters,
CPU base clock affected multi-threaded execution most with a linear correlation of \baseFreqMulTCor,
whereas boost frequency was the most influential parameter for single-threaded runs (correlation coefficient \boostFreqOneTCor);
cache size played no significant role.
When comparing our benchmark scores with popular high-performance computing benchmarks,
we found a strong linear relationship with Dolfyn
($R^2=\num[round-mode=places,round-precision=2]{\dolfynOneTRSq}$)
in the single-threaded scenario.
Using data from the 3DMark CPU Profile consumer benchmark,
we created a linear model for optimal (multi-threaded) Isabelle performance.
When validating,
the model has an average $R^2$-score of \num[round-mode=places,round-precision=2]{\modelRSq};
the mean absolute error in the final model corresponds to a wall-clock time of \SI[round-mode=places,round-precision=1]{\modelFinalTimeMae}{\second}.
With a dataset of true median values for the 3DMark,
the error improves to \SI[round-mode=places,round-precision=1]{\modelPrivateFinalTimeMae}{\second}.
\end{abstract}


\maketitle

% !TeX root = ../main.tex

\section{Introduction}
Choosing appropriate hardware and tuning configuration parameters
is a common task when one wants to run software optimally.
For a complex and truly parallel interactive proof assistant such as Isabelle,
many factors influence run-time performance:
The prover needs a Java and a Meta Language (ML) run-time,
the number of threads is variable,
as is the amount of heap memory --
which in turn 
(in combination with the CPU architecture family)
dictates which ML platform and hence Poly/ML backend may be used.
On a hardware level, CPU specs, the memory hierarchy, and interconnects
all influence how well the software components perform and how the system as a whole behaves.
The parallel efficiency of Isabelle
(i.e., the ratio of actual time versus sequential time divided by the number of parallel units)
decays according to a non-linear characteristic~\cite{Parallel2009Wenzel}, as is the case in most parallel systems.
As a result, there is no single hardware or software characteristic that dominates the observed performance behavior.

In Isabelle,
performance is important both in the interactive mode (such that processing changes and running solvers is faster)
and in a batch build mode, where \emph{sessions} 
(i.e., collections of formalizations) can be processed.
Independent sessions can even be run in parallel with multiple ML processes.

However, making informed decisions on hardware is no trivial task.
Members of the Isabelle community have to rely on word of mouth to determine which processors and memory to use,
and configuration parameters
(such as the number of threads or heap sizes)
are largely folk knowledge
-- backed by experience collected over the years, ad-hoc experiments, and sometimes intuition.
While there is some performance data available,
it is not helpful in that regard as it only covers a very small number of machines.

% problem
With new and exciting hardware being developed at a fast pace,
one can often be overwhelmed by the sheer variety of hardware options available.
Hence, the question of which hardware to recommend for Isabelle can often not be answered exhaustively or satisfactory.
This is relevant both for individuals working with Isabelle,
and for the larger-scale server infrastructure maintained for continuous integration and for Isabelle and the Archive of Formal Proofs.

% solution
To alleviate this problem,
a solid data base with performance benchmark results
for a wide variety of involved hardware and configurations
is needed.
Not only would that directly answer the question of optimal configurations for a given system
and allow one to compare the hardware on the market,
but such a collection of data
(if large enough, and kept up to date)
would also allow one to predict performance of other hardware for which no Isabelle data is available yet.

% contribution
In this paper,
we outline our Isabelle community benchmark,
discuss the immediate results and findings,
and derive a model to predict the Isabelle performance of unknown CPUs
with the help of widely used benchmarks for which more data is retrievable.
Our source-code and data and is made available publicly\footnote{\texttt{2022-paper} folder in \url{https://isabelle.systems/benchmark}}.
% Organization
Section~\ref{sec:related} covers related work;
we explain our benchmark set-up in Section~\ref{sec:benchmark},
and discuss the results in Section~\ref{sec:results}.
In Section~\ref{sec:conclusion},
we conclude and discuss future work.

% !TeX root = ../main.tex

\section{Related Work}\label{sec:related}
Parallel run-time performance has been first analyzed for Isabelle
when parallelism was introduced by \citeauthor{Parallel2009Wenzel} in~\cite{Parallel2009Wenzel}.
Benchmarks for multiple different sessions on a single test machine already showed
that the speedup
(in terms of run-time)
peaked at three worker threads with a factor of \num{3.0},
and slightly decreased for four cores.
\citeauthor{PolyParallel2010Matthews} described the necessary adaptations to the Poly/ML run-time
that were necessary for introducing parallelism,
and analyzed the resulting bottlenecks~\cite{PolyParallel2010Matthews}.
They found that the parallelization model for Isabelle sometimes failed to fully utilize all worker threads.
Moreover, the synchronization model that uses a single signal across all threads for guarded access
was identified (but not analyzed) as a potential bottleneck.
Finally, it was observed that the single-threaded garbage collection is responsible for up to \SI{30}{\percent} CPU-time for \num{16} threads.
Overall, a maximum speedup of \num{5.0} to \num{6.2} could be achieved
using \num{8} threads.

In automatic theorem provers, run-time is an important factor,
since it can dictate whether a goal can be proven within the given cut-off time.
As a result, much research includes analysis of the run-time performance of provers
or individual prover components.
Typically, only a single hardware configuration is used,
which is reasonable for the analysis for single-threaded systems~\cite{PerformanceESat2016Schulz}.
However, since performing such analysis on a wide range of different hardware is often impractical,
run-time performance analysis of parallel approaches
is frequently carried out on single systems or clusters~\cite{PerformanceOR1991Ertel,ParallelDeduction1992Jindal,ParallelHyper2001Wu}.
These results don't always generalize, because the hardware used can have a significant impact on the observed results.

In contrast, results for the Isabelle \texttt{sledgehammer} proof-finder tool show that when running \emph{multiple} automatic provers to solve a goal,
run-time becomes less important:
In their \emph{judgement day} study~\cite{Judgementday2010Boehme},
\citeauthor{Judgementday2010Boehme} found that running three different Automated Theorem Provers for five seconds each
solved as many goals as running the most effective one for \SI{120}{\second}.
Subsequently, in direct follow-up work~\cite{SMTHammer2011Blanchette}, run-time was not analyzed.

For automatic provers, a large range of benchmarks exist to judge their effectiveness on a given set of problems. 
%A large range of benchmarks exists to judge the effectiveness of automatic provers,
One of these is the widely known TPTP library~\cite{TPTP2009Sutcliffe}.
However, there is not much work investigating the effect of hardware in the field of automated reasoning.
To the best of our knowledge,
there exists no other benchmark comparing the hardware impact on run-time performance of any theorem prover,
and this is the first work that analyzes this effect on a wide range of different hardware.
% !TeX root = ../main.tex

\section{Benchmarking Methodology}\label{sec:benchmark}
% setup
The benchmark has to fulfill to multiple requirements:
It needs to capture typical computations found in Isabelle
--- mostly symbolic computation ---,
have a reasonable run-time for the end user,
and motivate users to want to see how their machines perform
(i.e., results should be self-evident).
We settled for a clean build of the HOL-Analysis session:
It is a typical Isabelle formalization
which runs in approximately five minutes on typical consumer machines.
Many Isabelle users have likely run a similar workload for their own work in the past.

While users can easily contribute results for their favourite Isabelle configuration,
we supplied a small script to run a comparable set of configurations automatically
\footnote{Documentation and code at \url{https://isabelle.systems/benchmark}}.
This way,
the whole benchmark can be run with a single command
(assuming a working installation of Isabelle 2021-1)
on any platform.
We vary the underlying ML platform between $64\_32$ ($64$-bit mode with $32$-bit values) and true \num{64}-bit mode,
heap sizes of both the ML and JVM process
(set to the same value to reduce the number of linear combinations,
as early benchmark results indicated they play only a minor role here),
and the number of worker threads for parallel proof checking.

Results are collected in a collaborative spreadsheet\footnote{\url{https://docs.google.com/spreadsheets/d/12GhEwSNSopowDBq5gSem3u39fliiIcoTIZHMnX4RE3A}}
with automatically updated figures for fastest CPUs and parallel efficiency.
% community
The benchmark is not intended as one-shot experiment,
but rather as a continuous community effort
to maintain an overview over the Isabelle computing landscape as new hardware emerges.
It is being kept open for new results, and will be maintained for future Isabelle versions.

\subsection{Benchmark Score (Isascore)}
For the benchmark results,
we use the wall-clock build time as an intuitive metric.
Together with the well-known HOL-Analysis build target,
the metric immediately gives a good understanding of Isabelle performance.
However, it is not well suited to compare to other metrics such as throughput,
because the relationship between time to solution and throughput is inverse.
To still allow using simple linear models such as the Pearson correlation,
we introduce a benchmark score that we call \emph{Isascore}.
It reflects the number of HOL-Analysis runs one could complete in a day, i.e.:
\begin{equation}
    \text{Isascore}=\frac{\SI{1}{\day}}{\text{wall-clock time}}
\end{equation}

\subsection{Threats to Validity}
The experiments discussed in this paper could not be performed in a controlled environment,
since they were run by members of the Isabelle community rather than exclusively by the authors of this paper.
This means that various outside factors may have influence on the reported results,
though it seems reasonable to assume that those factors should usually be constant between different configurations of the same benchmark run.
The effect of machine-local anomalies can be mitigated for hardware where we received several independent measurements by using statistical techniques.
Furthermore,
due to reasons of practicality in orchestrating data collection,
extended system specifics beyond the CPU model, OS, and memory configuration were not recorded.
There is a possibility that relevant parameters may have been missed.
Therefore, like all performance benchmarks, these results represent upper bounds of what might be achieved with a given system configuration.
Lastly,
while the benchmark was posted on the Isabelle-users mailing list,
in principle the data entry was open to public and could have been misused.
% !TeX root = ../main.tex

\section{Results}\label{sec:results}
At the time of writing this paper,
\num{\numRuns} results for a total of \num{\numConf} unique configurations have been reported,
utilizing \num{\numCpus} distinct CPUs.
Those include Intel Desktop/Server CPUs from Sandy Bridge to Alder Lake,
AMD Ryzen Zen2 to Zen4 processors as well as Epyc and Threadripper server systems,
a Fujitsu A64FX, and Apple M1 processors.

% !TeX root = ../main.tex

\DTLloaddb{topcpus}{top_cpus.csv}

\begin{table}[htbp]
  \caption{The five processors with the lowest time to solution, using each processor's optimal configuration (median value taken for multiple runs).}
  \label{tab:top_5_cpus}
  \begin{tabular}{lrrr}\toprule
    \textbf{Processor Model} & \textbf{Total Cores} & \textbf{Base Clock} & \textbf{Wall-Clock Time}
    \DTLforeach*{topcpus}{\cpu=cpu,\time=time,\cores=cores,\baseclock=baseclock}{
        \DTLiffirstrow{\\\cmidrule{1-4}}{\\}
        \cpu & \num{\cores} & \SI[round-mode=places,round-precision=1]{\baseclock}{\giga\hertz} & \SI{\time}{\second}
    }
    \\\bottomrule
  \end{tabular}
\end{table}

Table \ref{tab:top_5_cpus} shows the five CPUs with the lowest time to solution,
using the median value as an aggregate for multiple runs of the same configuration.
Older Intel and AMD consumer hardware is surpassed by the Apple M1 Pro chip;
only the most recent Intel core line performs better.
Due to the nature of the benchmark, server and high performance hardware does not rank highly,
with the best performing system (2x AMD Epyc 7742) clocking in at \SI{\bestServerTime}{\second}.

In the following, we analyze how Isabelle configuration influences performance,
investigate the impact of hardware parameters,
and then compare our results to other computational benchmarks.
Where individual CPUs were concerned,
we filtered out special system configurations (e.g., overclocked hardware, dual-cpu systems, power-saving mode).
We also encountered a small number of extreme outliers where Isabelle run-time was much longer than expected.
For two of those, we could identify the user and investigate;
in both cases, the system configuration was at fault
(excessive swapping, UEFI set to \enquote{silent mode})
and when corrected, results were much closer to the rest of the data.
We could not investigate the third extreme outlier but excluded it from the following,
since it is likely to stem from a similar cause.

\subsection{Multi-Threaded Performance}\label{sec:performance_threads}
\begin{figure}[htbp]
    \centering
    \begin{tikzpicture}
    \begin{groupplot}[
      group style={
        group name=timebythreads,
        group size=2 by 1,
        x descriptions at=edge bottom,
        y descriptions at=edge left,
        horizontal sep=0,
      },
      width=0.5\textwidth,
      height=192pt,
    ]
    \nextgroupplot[
      xmode=log,
      xtick={1,2,4,8,16,32,64,128},
      xlabel={Thread Count (log scale)},
      every axis x label/.append style={at=(ticklabel cs:1.0)},
      x label/.append style={at=(ticklabel cs:1.0)},
      log ticks with fixed point,
      ylabel={Wall-Clock Time},
      y unit={\si{\minute}},
      ytick distance=480,
      ymax=3540,
      ymin=0,
      yticklabel={\pgfmathparse{int(\tick/60)/100}\pgfmathprintnumber[skip 0.=true, dec sep={}, fixed,precision=2,zerofill]{\pgfmathresult}:\pgfmathprintnumber[skip 0.=true, dec sep={}, fixed,precision=2,zerofill]{0.0}},
    ]
    \foreach \cpu in \cpus {
    	\addplot[
    	  discard if not={cpu}{\cpu},
          color=black,
          mark=*,
          mark options={color=black,fill=white,scale=1,opacity=0.9},
          opacity=0.5
        ] table [
          col sep=comma,
          x=threads,
          y=time
        ] {time_by_threads.csv};
    }
    \nextgroupplot[
      xmode=log,
      xtick={1,2,4,8,16,32,64,128},
      log ticks with fixed point,
      ylabel={CPU Time per Thread},
      yticklabel pos=right,
      ylabel near ticks,
      ytick distance=480,
      y unit={\si{\minute}},
      ymax=3540,
      ymin=0,
      yticklabel={\pgfmathparse{int(\tick/60)/100}\pgfmathprintnumber[skip 0.=true, dec sep={}, fixed,precision=2,zerofill]{\pgfmathresult}:\pgfmathprintnumber[skip 0.=true, dec sep={}, fixed,precision=2,zerofill]{0.0}},
    ]
    \foreach \cpu in \cpus {
    	\addplot[
          discard if not={cpu}{\cpu},
          color=black,
          mark=*,
          mark options={color=black,fill=white,scale=1,opacity=0.9},
          opacity=0.5
        ] table [
          col sep=comma,
          x=threads,
          y=cputime
        ] {cputime_by_threads.csv};
    }
    \end{groupplot}
    \end{tikzpicture}
    \caption{Run-time by number of threads (log scale): Wall-clock time on the left, CPU time per thread on the right.}
    \label{fig:time_by_threads}
\end{figure}
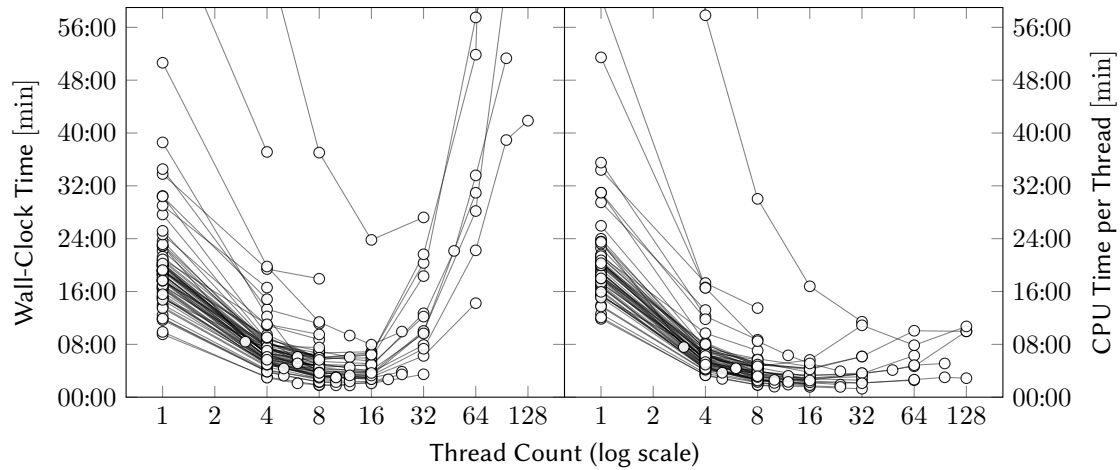

The number of threads used plays a major role in the overall performance.
\autoref{fig:time_by_threads} illustrates how the wall clock time and CPU time compare from a single thread to up to \num{128} threads.
The optimal wall-clock time is achieved with \numrange{8}{16} threads
depending on the hardware and greatly increases if more threads are used.
This is typical behavior for a strong-scaling benchmark like ours,
where the relative impact of communication increases with an increase in the number of threads used.
For more than the optimal number of threads,
the run-time increases substantially.
The underlying limitations of the parallel computing model 
-- the single-threaded garbage collection of the Poly/ML run-time and worker starvation after parallelization is saturated --
were already discussed in~\cite{PolyParallel2010Matthews},
albeit tests were run on a machine with 32 cores.
It might be a surprise that the scalability is so low when distributing across more threads.
In contrast,
the CPU time divided by number of threads
(which is not an ideal metric, but the only feasible solution due to the nature of the benchmark)
flattens out at eight threads.
In small-scale experiments, we found that the JVM process takes up a constant amount of CPU time independent of the number of threads (about \SI{26}{\percent} in single-core mode).
This means that there is not too much computation overhead
but the hardware can not be properly utilized by the ML process,
most likely due to the single-threaded garbage collection that stops all threads when running.
This is an inherently sequential task, which means that Amdahl's Law (the speedup is limited by $1 / (1-\text{parallelizable portion})$~\cite{amdahls}) limits the achievable speedup for this problem.
% !TeX root = ../main.tex

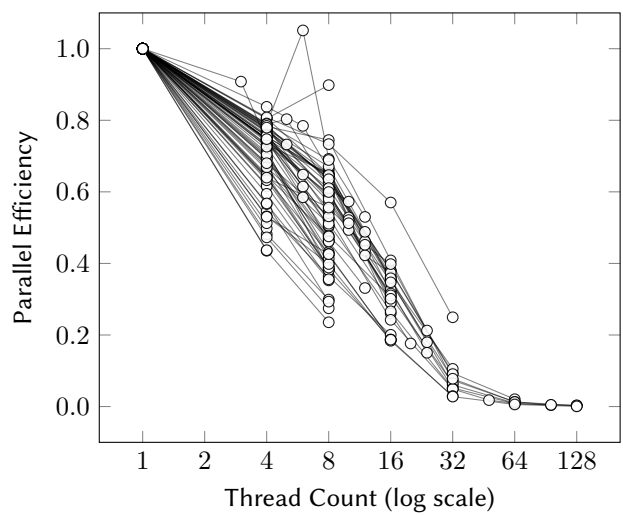
\begin{figure}[htbp]
    \centering
    \begin{tikzpicture}
    \begin{axis}[
      xmode=log,
      xtick={1,2,4,8,16,32,64,128},
      xlabel={Thread Count (log scale)},
      log ticks with fixed point,
      yticklabel={\pgfmathprintnumber[zerofill,precision=1,fixed]{\tick}},
      ymin=-0.1,
      ymax=1.1,
      ylabel={Parallel Efficiency},
    ]
    \foreach \cpu in \cpus {
    	\addplot[
    	  discard if not={cpu}{\cpu},
          color=black,
          mark=*,
          mark options={color=black,fill=white,scale=1,opacity=0.9},
          opacity=0.5
        ] table [
          col sep=comma,
          x=threads,
          y=rel
        ] {parallel_efficiency.csv};
    }
    \end{axis}
    \end{tikzpicture}
    \caption{Parallel efficiency by number of threads. The HOL-Analysis build does not scale past a low number of threads, with parallel efficiency already down by half when using eight threads.}
    \label{fig:parallel_efficiency}
\end{figure}

The parallel efficiency paints a similar picture in \autoref{fig:parallel_efficiency},
decreasing  almost linearly (on the logarithmic x-axis) up to \num{32} threads
at which it is at a median of \num[round-mode=places,round-precision=3]{\effLin}.
With the number of threads tending to the limit of \num{128},
it approaches \num[round-mode=places,round-precision=3]{\effEnd}.
There is an outlier where the parallel efficiency is over one --
super-linear speedup is unusual but can appear in practice because of caching effects or resource contention in the measured system.
% !TeX root = ../main.tex

\subsection{Performance Impact of Heap Memory}
As preliminary results indicated that heap memory
(as long as sufficient)
only plays a minor role in performance,
we keep the JVM and Poly/ML processes at the same heap size.
We know from experience that a few gigabytes of memory suffice for HOL-Analysis;
however, increased parallelism requires more memory in principle due to memory overhead.
Hence, the range of examined heap sizes depends on the number of threads used.
\autoref{fig:heap_boxplot} shows the change in run-time for different heap settings relative to the minimal setting.
The boxes capture the \num{25} and \num{75} percentiles as height and sampling size as width;
whiskers correspond to the extreme values.
The results show that performance is not affected very much by heap size.
Following the line of medians,
wall-clock time slightly increases above \SI{16}{\giga\byte}
(where the \SI{64}{\bit} Poly/ML backend needs to be used, as the more efficient $64\_32$ mode does not allow more than \SI{16}{\giga\byte}),
as well as for very large values.
We observed a single outlier for \num{64} threads and \SI{128}{\giga\byte} heap memory
at a relative factor of \num[round-mode=places,round-precision=2]{\heapOutlier}.
% !TeX root = ../main.tex
\makeatletter
\pgfplotsset{
    boxplot prepared from table/.code={
        \def\tikz@plot@handler{\pgfplotsplothandlerboxplotprepared}%
        \pgfplotsset{
            /pgfplots/boxplot prepared from table/.cd,
            #1,
        }
    },
    /pgfplots/boxplot prepared from table/.cd,
        table/.code={\pgfplotstablecopy{#1}\to\boxplot@datatable},
        row/.initial=0,
        make style readable from table/.style={
            #1/.code={
                \pgfplotstablegetelem{\pgfkeysvalueof{/pgfplots/boxplot prepared from table/row}}{##1}\of\boxplot@datatable
                \pgfplotsset{boxplot/#1/.expand once={\pgfplotsretval}}
            }
        },
        make style readable from table=lower whisker,
        make style readable from table=upper whisker,
        make style readable from table=lower quartile,
        make style readable from table=upper quartile,
        make style readable from table=median,
        make style readable from table=lower notch,
        make style readable from table=upper notch,
        make style readable from table=sample size,
        make style readable from table=draw position
}
\makeatother

\begin{figure}[htbp]
\centering
\begin{tikzpicture}
\pgfplotstableread[col sep = comma]{rel_time_by_heap_box.csv}\reltimebyheapbox
\begin{groupplot}[
  group style={
    group name=threads,
    group size=1 by 4,
    vertical sep=0pt,
    x descriptions at=edge bottom,
    y descriptions at=edge left,
  },
  height=120pt,
  width=307pt,
  xmin=2.5,
  xmax=9.5,
  ymin=0.7,
  ymax=1.3,
  xlabel={Heap Memory (log scale)},
  x unit={\si{\giga\byte}},
  boxplot/draw direction=y,
  boxplot/variable width,
  xticklabel={\pgfmathparse{pow(2,\tick)}\pgfmathprintnumber[int trunc]{\pgfmathresult}},
  boxplot/every whisker/.style={gray},
  boxplot/every median/.style={very thick},
  boxplot/draw/median/.code={},
]
\nextgroupplot
  \addplot[forget plot,boxplot prepared from table={table=\reltimebyheapbox,row=1,lower whisker=lw,upper whisker=uw,lower quartile=lq,upper quartile=uq,median=med,sample size=ss,draw position=heap}, boxplot prepared] coordinates {};
  \addplot[forget plot,boxplot prepared from table={table=\reltimebyheapbox,row=2,lower whisker=lw,upper whisker=uw,lower quartile=lq,upper quartile=uq,median=med,sample size=ss,draw position=heap}, boxplot prepared] coordinates {};
  \addplot[forget plot,boxplot prepared from table={table=\reltimebyheapbox,row=3,lower whisker=lw,upper whisker=uw,lower quartile=lq,upper quartile=uq,median=med,sample size=ss,draw position=heap}, boxplot prepared] coordinates {};
  \addplot[
    densely dotted,
    mark=-,
    mark size={3pt},
    mark options={solid,very thick},
    discard if not={threads}{8},
  ] table [
    col sep=comma,
    x=heap,
    y=med,
  ] {rel_time_by_heap_box.csv};
  \addlegendentry{medians};
\nextgroupplot
  \addplot[boxplot prepared from table={table=\reltimebyheapbox,row=5,lower whisker=lw,upper whisker=uw,lower quartile=lq,upper quartile=uq,median=med,sample size=ss,draw position=heap}, boxplot prepared] coordinates {};
  \addplot[boxplot prepared from table={table=\reltimebyheapbox,row=6,lower whisker=lw,upper whisker=uw,lower quartile=lq,upper quartile=uq,median=med,sample size=ss,draw position=heap}, boxplot prepared] coordinates {};
  \addplot[boxplot prepared from table={table=\reltimebyheapbox,row=7,lower whisker=lw,upper whisker=uw,lower quartile=lq,upper quartile=uq,median=med,sample size=ss,draw position=heap}, boxplot prepared] coordinates {};
  \addplot[boxplot prepared from table={table=\reltimebyheapbox,row=8,lower whisker=lw,upper whisker=uw,lower quartile=lq,upper quartile=uq,median=med,sample size=ss,draw position=heap}, boxplot prepared] coordinates {};
  \addplot[
    densely dotted,
    mark=-,
    mark size={3pt},
    mark options={solid,very thick},
    discard if not={threads}{16},
  ] table [
    col sep=comma,
    x=heap,
    y=med,
  ] {rel_time_by_heap_box.csv};
\nextgroupplot[ylabel={Relative Run-Time},every axis y label/.append style={at=(ticklabel cs:1.0)}]
  \addplot[boxplot prepared from table={table=\reltimebyheapbox,row=10,lower whisker=lw,upper whisker=uw,lower quartile=lq,upper quartile=uq,median=med,sample size=ss,draw position=heap}, boxplot prepared] coordinates {};
  \addplot[boxplot prepared from table={table=\reltimebyheapbox,row=11,lower whisker=lw,upper whisker=uw,lower quartile=lq,upper quartile=uq,median=med,sample size=ss,draw position=heap}, boxplot prepared] coordinates {};
  \addplot[boxplot prepared from table={table=\reltimebyheapbox,row=12,lower whisker=lw,upper whisker=uw,lower quartile=lq,upper quartile=uq,median=med,sample size=ss,draw position=heap}, boxplot prepared] coordinates {};
  \addplot[boxplot prepared from table={table=\reltimebyheapbox,row=13,lower whisker=lw,upper whisker=uw,lower quartile=lq,upper quartile=uq,median=med,sample size=ss,draw position=heap}, boxplot prepared] coordinates {};
  \addplot[boxplot prepared from table={table=\reltimebyheapbox,row=14,lower whisker=lw,upper whisker=uw,lower quartile=lq,upper quartile=uq,median=med,sample size=ss,draw position=heap}, boxplot prepared] coordinates {};
  \addplot[
    densely dotted,
    mark=-,
    mark size={3pt},
    mark options={solid,very thick},
    discard if not={threads}{32},
  ] table [
    col sep=comma,
    x=heap,
    y=med,
  ] {rel_time_by_heap_box.csv};
\nextgroupplot
  \addplot[boxplot prepared from table={table=\reltimebyheapbox,row=16,lower whisker=lw,upper whisker=uw,lower quartile=lq,upper quartile=uq,median=med,sample size=ss,draw position=heap}, boxplot prepared] coordinates {};
  \addplot[boxplot prepared from table={table=\reltimebyheapbox,row=17,lower whisker=lw,upper whisker=uw,lower quartile=lq,upper quartile=uq,median=med,sample size=ss,draw position=heap}, boxplot prepared] coordinates {};
  \addplot[
    densely dotted,
    mark=-,
    mark size={3pt},
    mark options={solid,very thick},
    discard if not={threads}{64},
  ] table [
    col sep=comma,
    x=heap,
    y=med,
  ] {rel_time_by_heap_box.csv};
\end{groupplot}
\node[right = 0.2cm of threads c1r1.east,anchor=north,rotate=90] {\num{8} Threads};
\node[right = 0.2cm of threads c1r2.east,anchor=north,rotate=90] {\num{16} Threads};
\node[right = 0.2cm of threads c1r3.east,anchor=north,rotate=90] {\num{32} Threads};
\node[right = 0.2cm of threads c1r4.east,anchor=north,rotate=90] {\num{64} Threads};\end{tikzpicture}
  \caption{Relative run-time changes to minimal memory configuration, by heap memory. The available heap memory has only a marginal impact on the run-time, with large amounts of available heap memory degrading performance independently of the number of threads used.}
  \label{fig:heap_boxplot}
\end{figure}
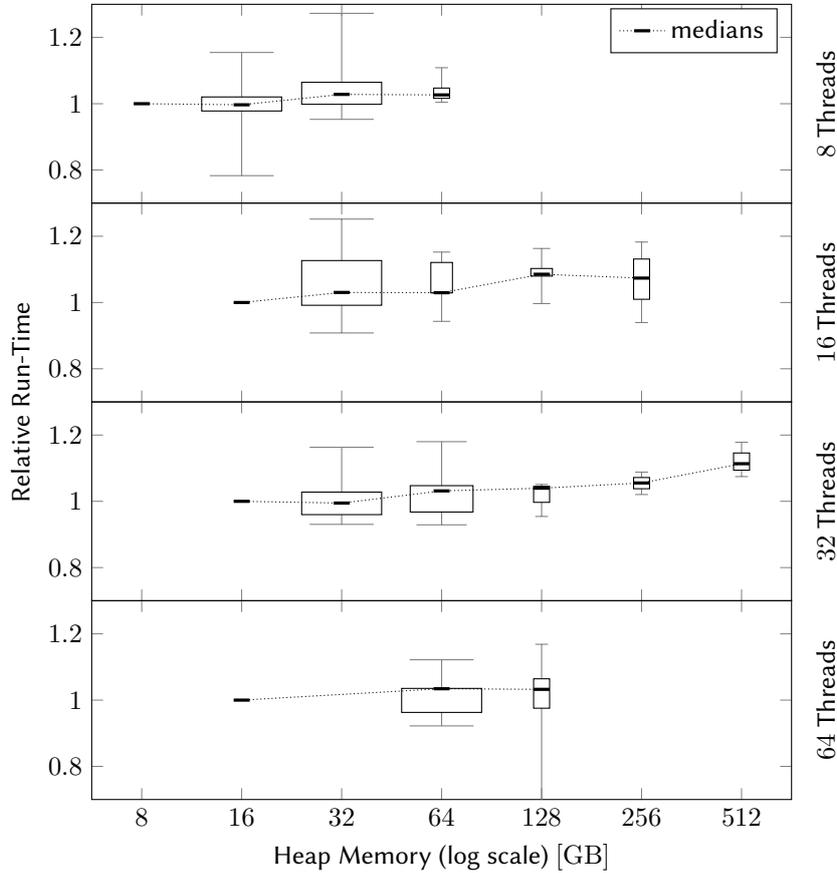

\subsection{Influence of Hardware Characteristics}
Based on folk knowledge about Isabelle performance,
we suspected that cache size would be a major factor;
it was debated whether boost clock speed would be relevant.
To test the hypotheses, we analyzed the impact of size of the L3-cache, base clock speed,
and maximal (boost) clock speed
(ignoring power-save cores where applicable)
on Isabelle performance.
Table \ref{tab:hardware_param_cor} shows the correlation between Isascore and those parameters (APA notation as explained in caption).
At our significance level of $0.05$,
we did not find cache size to impact performance significantly.
Base frequency is weakly correlated with the Isascore for a single thread (though at the edge of significance) and a bit more strongly (and much more significantly) in the multi-threaded scenarios.
Finally, boost frequency has a significant medium correlation for all modes,
which is strongest in the single-threaded configuration with a value of \num[round-mode=places,round-precision=2]{\cpuSpecsMulTCor}.
% !TeX root = ../main.tex

\DTLloaddb{cpuspecscor}{cpu_specs_cor.csv}

\begin{table}[htbp]
  \caption{Pearson correlation of hardware parameters with Isascore in APA notation: degrees of freedom 
(number of observations minus two),
$r$ (strength) and $p$ (significance). Cache size is left out as there is no significant correlation.}
  \label{tab:hardware_param_cor}
  \begin{tabular}{rll}\toprule
    \textbf{Mode} & \textbf{Base Frequency} & \textbf{Maximum Boost Frequency}
    \DTLforeach*{cpuspecscor}{\threads=threads,\corbase=corbase,\pbase=pbase,\nbase=nbase,\cormax=cormax,\pmax=pmax,\nmax=nmax}{
        \DTLiffirstrow{\\\cmidrule{1-3}}{\\}
        \threads & \ifdefempty\nbase{}{\cor{\nbase}{\corbase}{\pbase}} & \ifdefempty\nmax{}{\cor{\nmax}{\cormax}{\pmax}}
    }
    \\\bottomrule
  \end{tabular}
\end{table}

A possible explanation is that boost frequency can  only be sustained for a single core in most CPUs,
hence single-threaded performance profits from it a lot;
in the multi-threaded scenario, the actual core frequency is much closer to the base frequency
and thus its impact is larger.

\subsection{Comparison to Computational Benchmarks}
Performance benchmarks exist for many applications;
additionally, synthetic benchmarks are often used to evaluate hardware performance.
They can roughly be categorized into scientific computing versus consumer benchmarks. 
In the following, we compare the results of our Isabelle community benchmark with a number of publicly available datasets for such benchmarks.
For the comparison, we selected results with matching processors,
and matched the benchmarks' multi-thread setting (e.g., specific thread count, or all cores).
To obtain sufficiently large datasets,
we selected some of the most popular benchmarks.

\subsubsection{Benchmarks in High-Performance Computing}
The first analysis we wish to conduct is a comparison of Isabelle performance with some scientific programs.
For this analysis, we chose to import data from the High Performance Computing suite on OpenBenchmarking.
We selected the three benchmarks that had the most public results available
(in their primary configuration)
at the time of writing: \emph{Himeno}, \emph{NAMD}, and \emph{Dolfyn}.
Himeno\footnote{Results from \url{https://openbenchmarking.org/test/pts/himeno}} is an incompressible fluid analysis code written in Fortran and C~\cite{himeno}.
While a distributed memory parallel version exists (using MPI with Fortran VPP),
we concern ourselves with the sequential implementation.
NAMD\footnote{Results from \url{https://openbenchmarking.org/test/pts/namd}} is a shared memory parallel molecular dynamics code based on C++ and Charm++~\cite{namd}.
The data we use stems from machine-wide parallel trials.
Finally, Dolfyn\footnote{Results from \url{https://openbenchmarking.org/test/pts/dolfyn}} is a sequential computational fluid dynamics code based on Fortran~\cite{dolfyn}.

% !TeX root = ../main.tex

\begin{figure}[htbp]
\centering
\begin{tikzpicture}
\begin{groupplot}[
  group style={
    group name=hpcbenchmarks,
    group size=3 by 1,
    x descriptions at=edge bottom,
    y descriptions at=edge left,
    horizontal sep=55pt,
    every plot/.style={
    ymin=0,
    xmin=0,
    width=0.31\textwidth,
    height=192pt,
    }
  },
]

% himeno
\nextgroupplot[
   ylabel={Isascore (Single-threaded)},
   xlabel={MFLOPS},
   xmax=9900,
   xticklabel={\pgfmathprintnumber{\tick}},
   xtick distance=3000,
  ]
  \addplot[
    only marks,
    opacity=0.5,
  ] table [
    col sep=comma,
    x=score,
    y=isascore,
  ] {himeno_1t.csv};
  \addplot [
    thick,
  ] table [
    col sep=comma,
    x=score,
    y={create col/linear regression={x=score,y=isascore}},
  ] {himeno_1t.csv};
  \node[anchor=north east] at (rel axis cs:0.75,0.95) {$R^2=\num[round-mode=places,round-precision=2]{\himenoOneTRSq}$};
  
% dolfyn
\nextgroupplot[
    ylabel={Isabelle Time (Single-threaded)},
      y unit={\si{\minute}},
      ytick distance=720,
      ymin=0,
      yticklabel={\pgfmathparse{int(\tick/60)/100}\pgfmathprintnumber[skip 0.=true, dec sep={}, fixed,precision=2,zerofill]{\pgfmathresult}:\pgfmathprintnumber[skip 0.=true, dec sep={}, fixed,precision=2,zerofill]{0.0}},
        x unit={\si{\second}},
    xlabel={Execution Time},
    xticklabel={\pgfmathprintnumber{\tick}},
  ]
  \addplot[
    only marks,
    opacity=0.5,
  ] table [
    col sep=comma,
    x=score,
    y=time
  ] {dolfyn_1t.csv};
  \addplot [
    thick,
  ] table [
    col sep=comma,
    x=score,
    y={create col/linear regression={x=score,y=time}},
  ] {dolfyn_1t.csv};
  \node[anchor=north east] at (rel axis cs:0.75,0.95) {$R^2=\num[round-mode=places,round-precision=2]{\dolfynOneTRSq}$};
  
% NAMD vs mult-isabelle
\nextgroupplot[
ylabel={Isabelle Time (Multi-threaded)},
      y unit={\si{\minute}},
      ytick distance=360,
      ymin=0,
      yticklabel={\pgfmathparse{int(\tick/60)/100}\pgfmathprintnumber[skip 0.=true, dec sep={}, fixed,precision=2,zerofill]{\pgfmathresult}:\pgfmathprintnumber[skip 0.=true, dec sep={}, fixed,precision=2,zerofill]{0.0}},
    xlabel={Time per Simulated \si{\nano\second}},
    x unit=\si{\day},
  ]
\addplot[
    only marks,
    opacity=0.5,
  ] table [
    col sep=comma,
    x=score,
    y=time,
  ] {namd_mt.csv};
  \addplot [
    thick,
  ] table [
    col sep=comma,
    x=score,
    y={create col/linear regression={x=score,y=time}},
  ] {namd_mt.csv};
  \node[anchor=north east] at (rel axis cs:0.75,0.95) {$R^2=\num[round-mode=places,round-precision=2]{\namdMulTRSq}$};
\end{groupplot}
\node[above = 0.0cm of hpcbenchmarks c1r1.north] {Himeno};
\node[above = 0.0cm of hpcbenchmarks c2r1.north] {Dolfyn};
\node[above = 0.0cm of hpcbenchmarks c3r1.north] {NAMD};
\end{tikzpicture}
  \caption{Comparison of single-threaded (Himeno and Dolfyn) and multi-threaded (NAMD) Isabelle results with different high-performance computing benchmarks, with linear regression line.}
  \label{fig:hpc_benchmarks}
\end{figure}
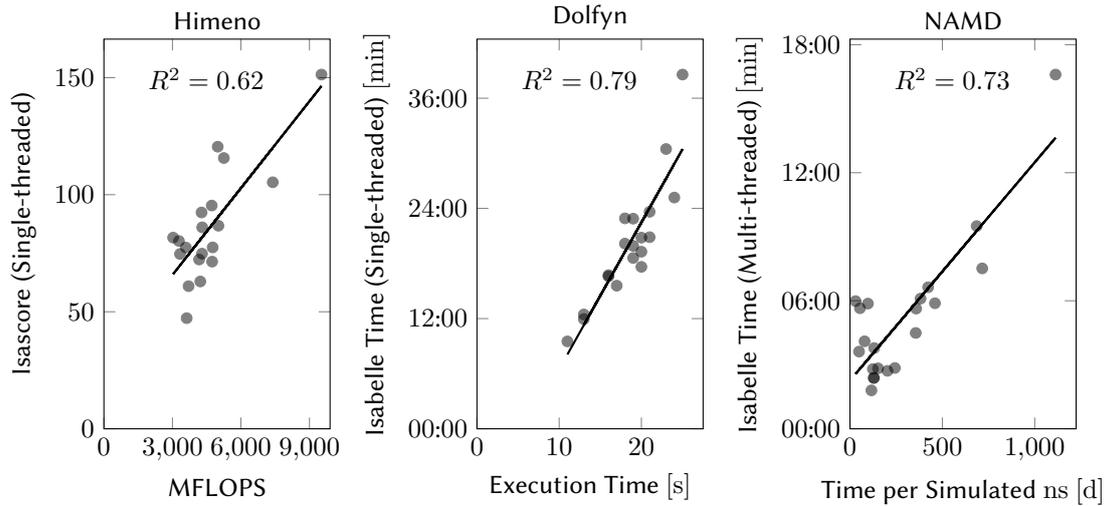

\autoref{fig:hpc_benchmarks} shows the results when correlating each of the high-performance computing benchmarks with Isabelle performance.
Himeno reports performance in terms of work done over time
(where higher is better),
while NAMD and Dolfyn measure time
(per simulated \si{\nano\second}, and to solution;
lower is better).
For Himeno, we therefore compare against Isascore,
while with NAMD and Dolfyn we compare against our observed wall clock time.

NAMD, as the only benchmark of these three
that scales well with parallel resources,
has no significant correlation with single-threaded Isabelle time.
However, it has a strong linear relation with multi-threaded time.
The two less scalable benchmarks correlate much closer with Isabelle single-thread performance, where Dolfyn has a particularly nice correlation that holds well for the most performant processors.
In both cases, correlation with multi-threaded Isabelle results is much worse
($R^2$-values: Himeno \num[round-mode=places,round-precision=2]{\himenoMulTRSq}, Dolfyn \num[round-mode=places,round-precision=2]{\dolfynMulTRSq}).
For both the Isabelle benchmark and Dolfyn,
the top processor that was tested is the same
(Intel i7-12700K),
and on both benchmarks it has a margin on the runner-ups.
This is also visible on the Himeno benchmark, where the 12700K produces the highest floating point throughput of all tested processors.
However, it is not a highly parallel processor, which is why its NAMD results are less favorable.
This again shows that Isabelle performance is significantly impacted by the single-thread performance of the underlying processor.

\subsubsection{Consumer CPU Benchmarks}
\begin{figure}[htbp]
\centering
\begin{tikzpicture}
\begin{groupplot}[
  group style={
    group name=benchmarks,
    group size=2 by 2,
    x descriptions at=edge bottom,
    y descriptions at=edge left,
    every plot/.style={
      width=209pt,
      height=192pt,
      ymin=0,
      ymax=170,
      xmin=-0.05,
      xmax=1.05,
    }
  },
]
\nextgroupplot
  \addplot[
    only marks,
    opacity=0.5,
    xlabel={Score},
  ] table [
    col sep=comma,
    x=score,
    y=isascore
  ] {cinebench_1t_rel.csv};
  \addplot [
    thick,
  ] table [
    col sep=comma,
    x=score,
    y={create col/linear regression={x=score,y=isascore}}
  ] {cinebench_1t_rel.csv};
  \node[anchor=north west] at (rel axis cs:0.05,0.95) {$R^2=\num[round-mode=places,round-precision=2]{\cinebenchOneTRSq}$};
\nextgroupplot
  \addplot[
    only marks,
    opacity=0.5,
  ] table [
    col sep=comma,
    x=score,
    y=isascore,
  ] {geekbench_1t_rel.csv};
  \addplot [
    thick,
  ] table [
    col sep=comma,
    x=score,
    y={create col/linear regression={x=score,y=isascore}}
  ] {geekbench_1t_rel.csv};
  \node[anchor=north west] at (rel axis cs:0.05,0.95) {$R^2=\num[round-mode=places,round-precision=2]{\geekbenchOneTRSq}$};
\nextgroupplot[
    xlabel={Normalized Score},
    every axis x label/.append style={at=(ticklabel cs:1.1)},
    ylabel={Isascore},
    every axis y label/.append style={at=(ticklabel cs:1.1)},
  ]
  \addplot[
    only marks,
    opacity=0.5,
  ] table [
    col sep=comma,
    x=score,
    y=isascore,
  ] {passmark_1t_rel.csv};
  \addplot [
    thick,
  ] table [
    col sep=comma,
    x=score,
    y={create col/linear regression={x=score,y=isascore}}
  ] {passmark_1t_rel.csv};
  \node[anchor=north west] at (rel axis cs:0.05,0.95) {$R^2=\num[round-mode=places,round-precision=2]{\passmarkOneTRSq}$};
\nextgroupplot
  \addplot[
    only marks,
    opacity=0.5,
  ] table [
    col sep=comma,
    x=score,
    y=isascore
  ] {x3dmark_1t_rel.csv};
  \addplot [
    thick,
  ] table [
    col sep=comma,
    x=score,
    y={create col/linear regression={x=score,y=isascore}}
  ] {x3dmark_1t_rel.csv};
  \node[anchor=north west] at (rel axis cs:0.05,0.95) {$R^2=\num[round-mode=places,round-precision=2]{\xdmarkOneTRSq}$};
\end{groupplot}
\node[above = 0.0cm of benchmarks c1r1.north] {Cinebench R15};
\node[above = 0.0cm of benchmarks c2r1.north] {Geekbench 5};
\node[above = 0.0cm of benchmarks c1r2.north] {PassMark CPU Mark};
\node[above = 0.0cm of benchmarks c2r2.north] {3DMark CPU Profile};
\end{tikzpicture}
  \caption{Comparison of single-threaded results with different consumer benchmarks, with linear regression line. Benchmark scores are normalized to $[0,1]$ range.}
  \label{fig:consumer_benchmarks}
\end{figure}
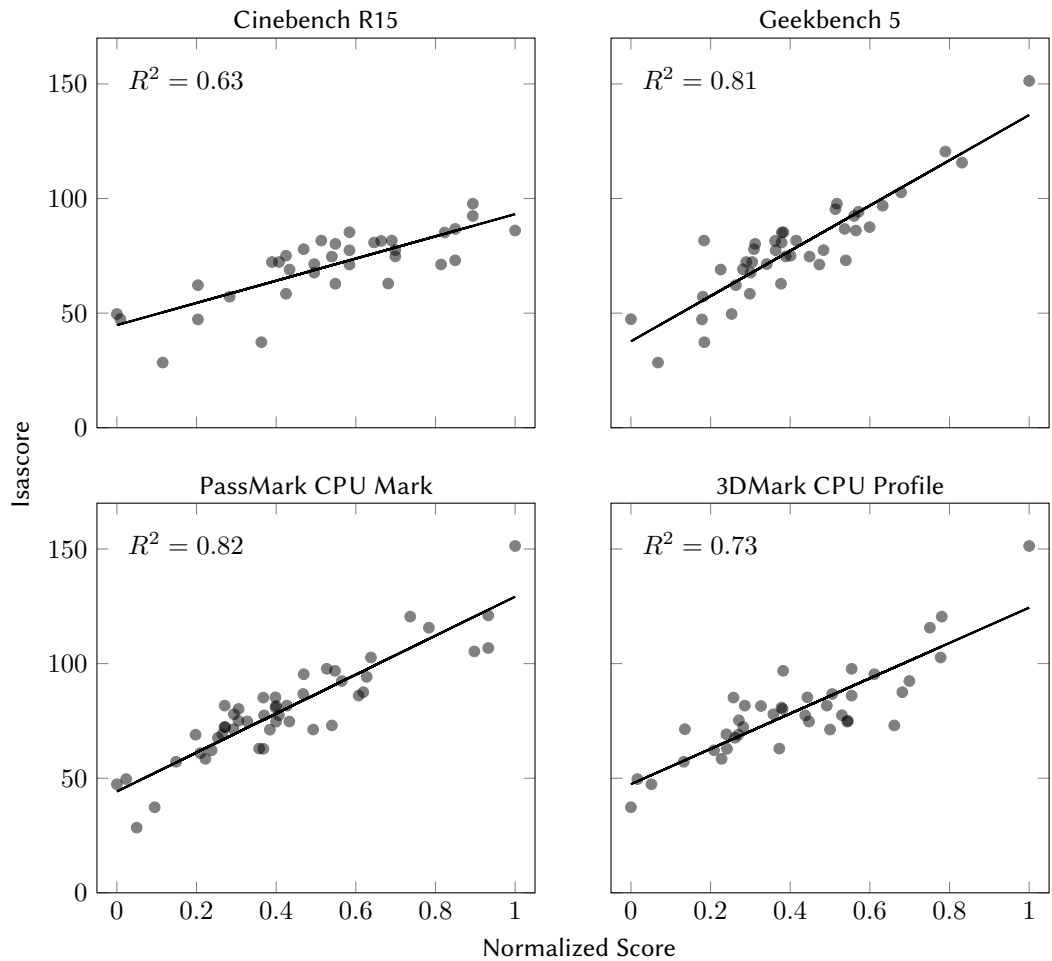

For our second comparison,
we chose some of the most common consumer benchmarks to compare to:
\emph{PassMark CPU Mark}\footnote{Results from  \url{https://www.cpubenchmark.net/CPU_mega_page.html}}, \emph{Geekbench 5}\footnote{Results from \url{https://browser.geekbench.com/processor-benchmarks.json}}, \emph{Cinebench R15}\footnote{Results from \url{https://us.rebusfarm.net/en/tempbench}}, and \emph{3DMark CPU Profile}\footnote{Results from \url{https://www.3dmark.com/search}, median over the top-\num{100} values}.
For sequential performance,
\autoref{fig:consumer_benchmarks} shows the scatter plots of Isascore to consumer benchmark scores,
which are normalized to a $[0;1]$ range so the plots can be compared against.
A strong positive relationship can be observed for all benchmarks,
with $R^2$-values in the range \numrange[round-mode=places,round-precision=2]{\cinebenchOneTRSq}{\passmarkOneTRSq}.
A few moderate outliers are present (possibly due to system configuration).
All in all, the Isabelle benchmark seems quite similar to those consumer benchmarks for a single thread.
% !TeX root = ../main.tex

This gives rise to prediction of Isabelle performance,
which would allow one to judge hardware on which Isabelle was not executed before.
However, single-threaded results are not meaningful for real-world performance,
and scaling them according to the average parallel efficiency did not yield helpful results
($R^2$-values: Cinebench \num[round-mode=places,round-precision=2]{\cinebenchMulTRSq}, Geekbench \num[round-mode=places,round-precision=2]{\geekbenchMulTRSq}, PassMark \num[round-mode=places,round-precision=2]{\passmarkMulTRSq}, 3DMark \num[round-mode=places,round-precision=2]{\xdmarkMulTRSq}).
Not many datasets for consumer benchmarks report on results for different number of threads,
most report only a single \enquote{multi-core} value where all threads are utilized.
An exception to that is the 3DMark CPU Profile benchmark,
where results are reported for \numrange{1}{16} threads individually
(in steps by power of two).
This allows us to create a better correlation, because all consumer benchmarks tested had a far better parallel efficiency in the limit
and were hence not suited for direct prediction.

\begin{figure}[htbp]
\centering
\begin{tikzpicture}
\begin{groupplot}[
  group style={
    group name=threads,
    group size=2 by 1,
    x descriptions at=edge bottom,
    y descriptions at=edge left,
    horizontal sep=56pt,
  },
  width=0.48\textwidth,
  height=192pt,
]
  \nextgroupplot[
  xlabel={Averaged \num{8}/\num{16}-Thread 3DMark CPU Profile Score},
  every axis x label/.append style={at=(ticklabel cs:1.0)},
  ylabel={Isascore},
  ymin=0,
  xmin=0,
  xmax=9900,
  legend style={at={(0.98,0.02)},anchor=south east},
  domain=0:11000,
  ]
  \addplot[
    only marks,
    opacity=0.5,
    forget plot
  ] table [
    col sep=comma,
    x=score,
    y=isascore,
  ] {x3dmark_mt.csv};
  \addplot[thick] {\modelAX*x+\modelB};
  \node[anchor=north west] at (rel axis cs:0.05,0.95) {$R^2=\num[round-mode=places,round-precision=2]{\modelFinalRSq}$};
  \addlegendentry{%
$\num[round-mode=places,round-precision=1]{\modelB}+\num[round-mode=places,round-precision=3]{\modelAX}\cdot x$};
  \nextgroupplot[
    ylabel={Isabelle Time},
    ytick distance=180,
    y unit=\si{\second},
    ymin=0,
    ymax=1080,
    yticklabel={\pgfmathparse{int(\tick/60)/100}\pgfmathprintnumber[skip 0.=true, dec sep={}, fixed,precision=2,zerofill]{\pgfmathresult}:\pgfmathprintnumber[skip 0.=true, dec sep={}, fixed,precision=2,zerofill]{0.0}},
    xmax=9900,
    xmin=0,
    domain=0:11000,
  ]
  \addplot[
    only marks,
    opacity=0.5,
  ] table [
    col sep=comma,
    x=score,
    y=time,
  ] {x3dmark_mt.csv};
  \node[anchor=north east] at (rel axis cs:0.95,0.95) {$\text{MAE}=\SI[round-mode=places,round-precision=1]{\modelFinalTimeMae}{\second}$};
  \addplot[thick, samples=1000] {(24*60*60) / ((\modelAX*x)+\modelB)};
\end{groupplot}
\end{tikzpicture}
  \caption{Scatter plots for final model of Isabelle performance. Linear model on Isascore on the left, corresponding wall clock run-times on the right.}
  \label{fig:predictor}
\end{figure}
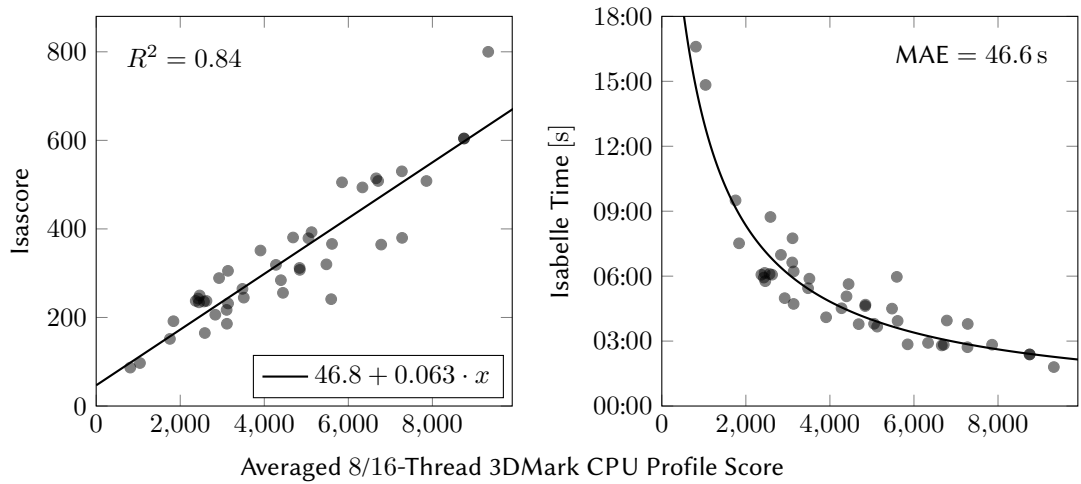

When using \num{8} and \num{16} threads in both the 3DMark and Isabelle benchmark,
score and Isascore are strongly to moderately correlated and have individual $R^2$-values of \num[round-mode=places,round-precision=2]{\xdmarkEightTRSq} and \num[round-mode=places,round-precision=2]{\xdmarkSixteenTRSq}, respectively.
This makes the 3DMark well suited for performance prediction.
Since the optimal number of threads is in between,
we use the average of its \num{8}-thread and \num{16}-thread results
to create a linear model for performance prediction
(tuning for a non-uniform split did not yield better results).
Using ten times ten-fold cross-validation
(i.e., averaging results over multiple iterations,
splitting the data into ten parts, and using each part as a test set and the remainder as training set),
the linear regression has an average $R^2$-value of \num[round-mode=places,round-precision=2]{\modelRSq}.
\autoref{fig:predictor} shows the final model ($R^2=\num[round-mode=places,round-precision=2]{\modelFinalRSq}$) and the resulting predictor for wall-clock time, which has a mean absolute error (MAE) of \SI[round-mode=places,round-precision=1]{\modelFinalTimeMae}{\second}.
However, that error is somewhat exaggerated by the data collection method:
The public 3DMark data only shows only the top-$100$ results, from which we use the median values.
The regression improves slightly when the
(non-public) true medians are used,
and the MAE decreases to \SI[round-mode=places,round-precision=1]{\modelPrivateFinalTimeMae}{\second}.
% !TeX root = ../main.tex

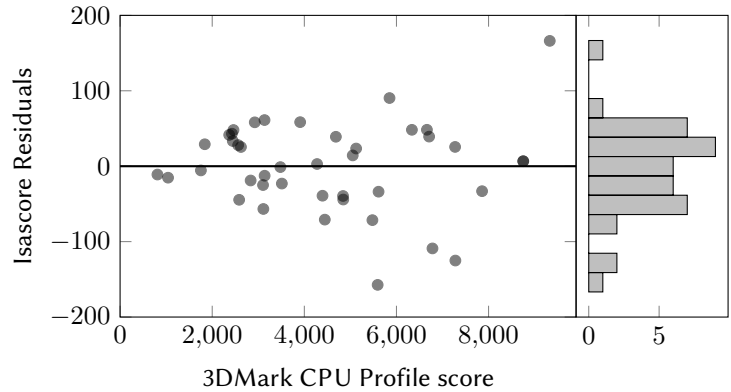
\begin{figure}[htbp]
\centering
\begin{tikzpicture}[
    /pgfplots/scale only axis,
    /pgfplots/width=6cm,
    /pgfplots/height=4cm
]
\begin{axis}[
  name=main axis,
  ylabel={Isascore Residuals},
  xlabel={3DMark CPU Profile score},
  xmin=0,
  xmax=9900,
  ymax=200,
  ymin=-200,
]
  \addplot[
    only marks,
    opacity=0.5,
  ] table [
    col sep=comma,
    x=score,
    y=residual,
  ] {x3dmark_mt_residuals.csv};
  \draw[thick] (rel axis cs:0,0.5) -- (rel axis cs:1,0.5);
\end{axis}

\begin{axis}[
    anchor=north west,
    at=(main axis.north east),
    width=2cm,
    ytick=\empty
]
\addplot [
    hist={bins=13,handler/.style={xbar interval}},
    x filter/.code=\pgfmathparse{rawy},
    y filter/.code=\pgfmathparse{rawx},
    fill=gray!50,
] table [col sep=comma] {x3dmark_mt_residuals.csv};
\end{axis}
\end{tikzpicture}
  \caption{Residual plot for final linear model, with histogram. The residuals show no apparent patterns and roughly follow a normal distribution.}
  \label{fig:predictor_residual}
\end{figure}

The residual plot displayed in \autoref{fig:predictor_residual} has no noticeable patterns
and the residual distribution roughly follows a normal distribution.
All in all, the model simplicity and good fit
indicate that this linear model is quite well suited for performance prediction,
as long as the other system parameters are kept within reasonable bounds
and the configuration is well tuned.
% !TeX root = ../main.tex

\section{Conclusion}\label{sec:conclusion}
This work resolves our questions on Isabelle performance for the 2021-1 version.
The Isabelle community benchmark that we initiated saw lively participation
and hundreds of results were reported for a total of \num{\numCpus} distinct CPUs.
The results form a solid data base for tuning of Isabelle configuration;
when not constrained, the optimal configuration is at \numrange{8}{16} threads with \SI{16}{\giga\byte} heap memory for both the Java and ML process
(at least for HOL-Analysis, larger sessions might require more).

When buying new hardware,
the benchmark results give a good indication of which processor is desirable for Isabelle.
Individual CPU parameters are not as important
as clock speeds are only correlated with medium strength (though boost clock more than base clock)
and cache size not significantly at all.
Instead, for hardware that has not yet been tested with Isabelle,
other benchmarks can greatly help in judging performance:
While the single-threaded Isabelle benchmark score is strongly correlated with many benchmarks
(most strongly with the Dolfyn high-performance benchmark and the PassMark CPU Mark),
the multi-threaded scenario was more difficult to model;
In the end, we found a good predictor by using 3DMark CPU Profile scores from \num{8} and \num{16} threads,
with a final mean absolute error of \SI[round-mode=places,round-precision=2]{\modelFinalTimeMae}{\second}.
The model has a good fit, and one can assume that it is fairly future-proof given that hardware from the last ten years is properly predicted.

\section{Future Work}\label{sec:future}

If the Isabelle computation model does not change,
there is not much left to be done on the topic of performance prediction:
The benchmark from which Isabelle performance is predicted is widely popular and data for new hardware is usually quickly added,
and the fit is about as good as one can hope for.
Still, the model should be validated after a few years,
and we are curious to see whether future hardware characteristics will be able to break the trend.

One other aspect that we did not touch on is running multiple independent sessions in parallel,
which is often possible on automated large-scale Isabelle builds, e.g., for the Archive of Formal Proofs.
This can be done on multiple processes that run independently of each other and greatly increases the usefulness of larger server CPUs with many cores;
then, other parameters such as memory bandwidth might be more important and have to be analyzed.
However, given the large cost of such machines,
it would be much more economical to instead distribute the build on multiple cheap but fast desktop CPUs
(especially when latency is a concern, not throughput).

\begin{acknowledgments}
Thanks to all the countless individuals who participated in the benchmark. Parts of the performance results have been obtained on systems in the test environment BEAST (Bavarian Energy Architecture \& Software Testbed) at the Leibniz Supercomputing Centre. NAMD was developed by the Theoretical and Computational Biophysics Group in the Beckman Institute for Advanced Science and Technology at the University of Illinois at Urbana-Champaign.
Many thanks to Phoronix Media (OpenBenchmarking), PassMark Software (PassMark), RebusFarm GmbH (Cinebench), Primate Labs Inc. (Geekbench) and especially UL Benchmarks (3DMark), for providing their benchmark data.
\end{acknowledgments}

\Urlmuskip=0mu plus 1mu\relax
\bibliography{bibliography}

\end{document}